\documentclass[fleqn,twoside]{article}
\usepackage{espcrc2}
\usepackage{graphicx}
\usepackage[figuresright]{rotating}

\title{Positron Annihilations at the Galactic Center:\\Generating More Questions Than Answers}

\author{Hasan Y\"{u}ksel\\\vspace*{0.1cm}{Department of Physics, Ohio State University, Columbus, Ohio 43210, USA}}
     
\begin{document}

\begin{abstract}
The bulge of our Galaxy is illuminated by the 0.511 MeV gamma-ray line flux from annihilations of nonrelativistic positrons. The emission is strongly concentrated at the Galactic Center (GC), in contrast to gamma-ray maps tracing nucleosynthesis (e.g., the 1.809 MeV line from decaying $^{26}$Al) or cosmic ray processes (e.g., the 1--30 MeV continuum), which reveal a bright disk with a much less prominent central region. If positrons are generated at relativistic energies, higher-energy gamma rays will also be produced from inflight annihilation of positrons on ambient electrons.  The comparison of the gamma-ray spectrum from inflight annihilation to the observed diffuse Galactic gamma-ray data constrains the injection energies of Galactic positrons to be less than $3$ MeV.
\end{abstract}

\maketitle

%
The 0.511 MeV gamma-ray line emission from the Galactic Center was the first gamma-ray line detected from outside our solar system. Now the recent SPI/INTEGRAL detected that $\sim 10^{50}$ positrons are annihilated every year~\cite{INTEGRALearly,Churazov04,Knodlseder05,Jean05,Weidenspointner06,Guessoum05,Strong05}, producing a 0.511 MeV gamma-ray line flux of $\left(1.07 \pm 0.03 \right) \times 10^{-3}$ photons cm$^{-2}$ s$^{-1}$. The SPI/INTEGRAL instrument also mapped the angular distribution of this emission, revealing a 2-dimensional Gaussian of $\simeq 8^\circ$ FWHM aligned with the GC. The bulge to disk ratio of the emission signal is much larger than observed in any other energy band~\cite{Strong98,Knodlseder99}. Many astrophysical~\cite{Models-Astro} and exotic~\cite{Models-Exotic} models were suggested to explain the flux and angular distribution of the 0.511 MeV radiation. 

Their injection energy is central to resolve the origin of the positrons. Positrons annihilating after energy loss, produce only gamma rays {\it at or below} 0.511 MeV, obscuring their true injection energies. We get around this by considering gamma rays produced by the Inflight Annihilation (IA) of energetic positrons with electrons in the interstellar medium. The IA radiation has already been used to probe astrophysical positrons in many settings, e.g., Refs.~\cite{Stecker,Murphy,Svensson,Aharonian88,Moskalenko}. We determine constraints on the initial injection energy by normalizing the intensity and angular distribution of the continuum flux from positron annihilation in flight to INTEGRAL and COMPTEL observations of diffuse Galactic gamma-rays.

%
We calculate the survival probability for injected positrons as follows (see Ref.~\cite{Beacom:2005qv} for details). 
We adopt monoenergetic injection for positrons at a total energy $E_0$. They remain confined to the GC by magnetic fields while they lose energy due to ionization. We calculate the Inflight Annihilation signal for a neutral medium since direct astrophysical probes suggest a weakly ionized medium at the GC. The energy loss rate for a positron due to collisions with electrons in a neutral hydrogen medium of number density $n_{\rm H}$ is~\cite{energyloss} $ \left|\frac{dE}{dx} \right| \simeq \frac{7.6 \times 10^{-26}}{\beta^2} \frac{n_{\rm H}}{0.1 \;\rm{cm}^{-3}} (\ln \gamma +6.6) \; \frac{\rm MeV}{\rm cm}, $ where $E$ is the positron energy, $\gamma = E/m_e $ is the Lorentz factor and $\beta$ is the velocity.  The fraction of positrons annihilating as they lose an energy $dE$ and travel a distance $dx$ is $ \frac{dN_{}(E)}{N_{}(E)}= n_{\rm H} \sigma(E)  \frac{dE}{| dE/dx |}, $ where $\sigma(E)$ is the annihilation cross section of positrons. Then the survival probability of positrons is
\begin{equation}
{\rm P}_{E_0\rightarrow E} =
\exp\left( - n_{\rm H}\int_{E}^{E_0} \sigma( E')\, \frac{ d E'} {| dE'/dx |}\right),
\label{eq:survival}
\end{equation}
as they lose energy from $E_{\rm 0}$ to $E$. The fraction of positrons that are annihilated is $\simeq$ 11 (5.5, 1.4) \% for injection energies of 10 (3, 1) MeV. The rate of nonrelativistic positron annihilation at the GC is $\dot N_{}(m_e) \sim  10^{50}$ year$^{-1}$. The small IA fraction increases the required positron injection rate to $\dot N_{}(E_0) = \dot N_{}(m_e)/ {\rm P}$, where ${\rm P} = {\rm P}_{E_0\rightarrow m_e}$ is terminal survival probability.

Nonrelativistic positrons may either directly annihilate with an electron, producing two 0.511 MeV gamma rays, or form a positronium bound state with an electron~\cite{Stecker,Leventhal} which annihilates to two gamma rays (each 0.511 MeV) 25\% of the time, and to three gamma rays (each less than 0.511 MeV) 75\% of the time. The relative intensities of the three-gamma continuum and two-gamma line emission at the GC fix the  positronium fraction to be $f = 0.967 \pm 0.022$~\cite{Jean05} so the true annihilation rate is 3.6 times {\it larger} than would be deduced from the 0.511 MeV flux alone. The IA of energetic positrons produces two gamma rays, and so the ratio of the total flux of IA to 0.511 MeV gamma rays is
\begin{equation}
\frac{\Phi_{\rm IA}}{\Phi_{0.511}}
=\frac{2 \;(1-{\rm P})}{2 \;(1 - \frac{3 f} {4}) \; {\rm P} }
=\frac{1}{1 - \frac{3 f} {4}}\; \frac{1-{\rm P}}{{\rm P}}.
\label{eq:ratios}
\end{equation}

\begin{figure}[t]
\includegraphics[width=2.75in,clip=true]{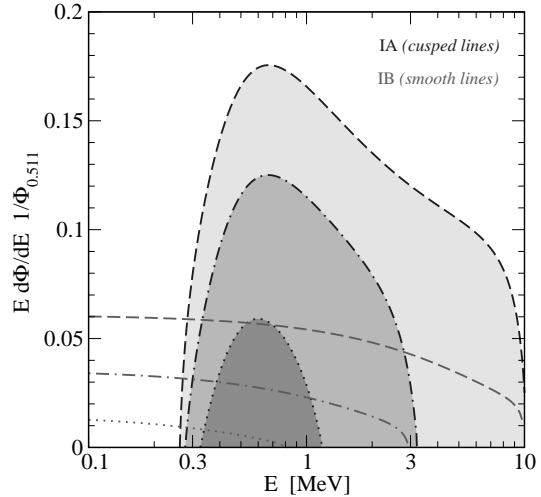}
\caption{Gamma-ray spectra (peaked set, inflight annihilation; flatter set, internal bremsstrahlung) from relativistic positrons, normalized to the 0.511 MeV flux; the injection energies are 1, 3, and 10~MeV (dotted, dot-dashed and dashed). Taken from Ref.~\cite{Beacom:2005qv}.
\label{fig:galpos-1}}
\end{figure}

In Fig.~\ref{fig:galpos-1}, we show the IA gamma ray spectra for various positron injection energies.
The shape of the integrated gamma-ray spectrum produced by the IA of positrons as they lose energy is:
\begin{equation}
\frac{d\Phi_{\rm IA}}{dk} = \frac{\Phi_{\rm 0.511}}{1 - \frac{3 f} {4}} \; \frac{n_{\rm H}}{\rm P}  
\int^{E_0}
{\rm P}_{E_0\rightarrow E} \; \frac{1}{2} \frac{d\sigma}{dk} \; \frac{dE}{| dE/dx |}.
\label{eq:IAspectrum}
\end{equation}
where $\frac{d\sigma}{dk}$ is the angle-averaged differential cross section~\cite{Stecker,Svensson,Aharonian00} for IA, in terms of the scaled gamma-ray energy $k = E_\gamma / m_e$ (weighted with the gamma-ray multiplicity of 2). The gamma rays above 0.511 MeV are spread over a broad energy range, but their detectability is greatly enhanced since the Galactic diffuse gamma-ray background is steeply falling with energy. 

\begin{figure}[t]
\includegraphics[width=2.75in,clip=true]{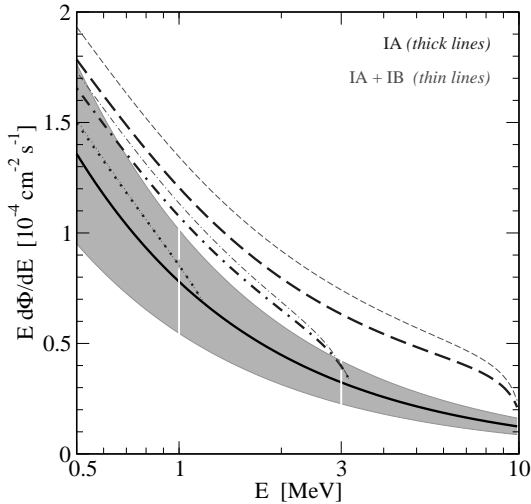}
\caption{The diffuse gamma ray flux is shown with a  solid line with  $\pm$30\% uncertainties at $5^{\circ}$-diameter region at the Galactic Center.  For positron injection energies of 1, 3, and 10 MeV (dotted, dot-dashed and dashed lines), the thick lines show how this would be {\it increased} by the inflight annihilations (thin lines also include the internal bremsstrahlung). Taken from Ref.~\cite{Beacom:2005qv}.
\label{fig:galpos-2}}
\end{figure}

%
Since the 0.511 MeV and IA gamma rays are both emitted isotropically, their angular distributions should be the same on average. We will normalize our subsequent results to the observed rate and angular distribution of the 0.511 MeV line flux.
The gamma rays at the GC will have two components; the calculated gamma-ray spectrum from IA (and Internal Breamstrahlung, IB) for an assumed injection energy and the diffuse gamma-ray background. The diffuse flux is slowly varying (showing no excess at the GC like that for the 0.511 MeV line), and we normalize it from the measured INTEGRAL and COMPTEL data averaged in the Galactic Plane. 

For the diffuse flux, we used the power law $ \frac{d\Phi}{dE} = \left( \frac{E}{0.09  {\rm MeV}} \right)^{-1.8} {\rm cm^{-2}\ s^{-1}\ sr^{-1}\ MeV^{-1}}, $ which reproduces the COMPTEL and  INTEGRAL measurements of diffuse flux at the inner galactic disk~\cite{Strong05,Strong98}. In Fig.~\ref{fig:galpos-2}, we show the diffuse spectrum, obtained by scaling the data averaged over a large region with the solid angle of $5^{\circ}$-diameter circle at the GC; the generous $\pm$30\% uncertainties are shown as a shaded band (primarily a systematic uncertainty on the normalization, due to subtracting detector backgrounds).
We also show how our predictions for the IA flux would increase the average diffuse flux in the same GC circle. If the positron injection energy is large, say 10 MeV or greater, the inflight annihilation would more than double the diffuse flux compared to the average diffuse flux. Since the observed diffuse flux has no prominent concentration at the GC, such high injection energies are clearly disallowed.
 
This argument can be strengthened by using the COMPTEL diffuse skymaps, and comparing to the measured flux in strips of $5^\circ$ longitude and $10^\circ$ latitude~\cite{Strong98}.  The 1--3 and 3--10 MeV skymaps both show a moderate dip and peak structure straddling the GC, suggesting that its origin is in the diffuse gamma-ray background, and unrelated to positrons.  If positrons are injected at 3 MeV, then the IA flux in the dip region at 1--3 MeV is comparable to the whole diffuse flux there. To be conservative (bearing in mind the uncertainties of such skymaps), even though this looks implausible, we assume it is possible, and thus determine our upper limit of 3 MeV for the positron injection energy.  At higher injection energies, there would also be a too-large inflight annihilation flux in the 3-10 MeV data.

Ref.~\cite{Sizun:2006uh} investigated the dependence of this limit on the ionized fraction of the ISM and the angular distribution of the annihilation signal in the context of light dark matter particles. They arrive at an upper bound of 3--7.5~MeV, which is still below 10 MeV lower limit of Ref.~\cite{Fayet:2006sa} from supernova cooling time scale, leaving no room for light dark matter particles with annihilation and scattering cross sections larger than neutrinos. 

%
We have considered constraints imposed on the injection energy of positrons at the Galactic Center by comparing their Inflight Annihilation gamma rays to the diffuse flux as measured by COMPTEL. The  recent high-quality data from INTEGRAL on the flux and angular distribution of the 0.511~MeV gamma-ray line enable a model-independent test on the mysterious origins of the Galactic positrons. While positron annihilations (and any possible inflight annihilation gamma rays at higher energies) are concentrated at the GC, the diffuse gamma rays are spread over the Galactic Plane, clearly pointing out distinct origins of two signals. Our constraints on injection energies are based on the fact that the predicted inflight annihilation gamma ray flux should not be too large compared to the measured  diffuse gamma flux. Our results are model independent and directly probe the positron injection energy, requiring it to be $< 3$ MeV which improves the previous limit of 20 MeV~\cite{Beacom04} by nearly an order of magnitude. 

Injection energies as small as 3 MeV are very close to the energy scale of nuclear beta decays from fresh nucleosynthesis products.  The angular map of positron injection positions given by IA and IB emission should more faithfully reveal the sources, in addition to the information from the injection energy scale alone.

I thank John Beacom for collaboration.  This work is supported by The Ohio State University and by NSF CAREER grant No.~PHY-0547102 to JB.


\end{document}